\documentclass[prb,twocolumn,showpacs]{revtex4}
\usepackage{graphicx}

\begin{document}
\title{High-pressure phase relations of Bi$_{2}$Sr$_{2}$CaCu$_{2}$O$_{8+\delta}$ single crystals }
\author{Xiao-Jia Chen, Viktor V. Struzhkin, Russell J. Hemley, and Ho-kwang Mao }
\affiliation{Geophysical Laboratory, Carnegie Institution of Washington, Washington, DC 20015}
\author{Chris Kendziora}
\affiliation{Code 6365, Naval Research Laboratory, Washington, DC 20375}
\date{\today}

\begin{abstract}
We have investigated the pressure dependence of the superconducting transition temperature $T_{c}$ up to 18 GPa
of Bi$_{2}$Sr$_{2}$CaCu$_{2}$O$_{8+\delta}$ single crystals ranging from the highly underdoped through the
nearly optimally doped to the highly overdoped level. For all three samples studied, $T_{c}$ is found to
increase initially and then saturate at a critical pressure $P_{c}$ but decrease modestly with further
compression. Oxygen doping tends to reduce the increase in $T_{c}$ and $P_{c}$. A new high-pressure phase 
diagram between the saturated $T_{c}$ and $P_{c}$ is then obtained. Theoretical interpretation is given by 
using the competition between the hole carrier density and pairing interaction strength based on the 
high-pressure transport data of the resistivity and Hall coefficient in this system.
\end{abstract}

\maketitle

\section{INTRODUCTION}

The doping dependence of the high-temperature superconductors (HTSCs) has become a significant issue due to
pseudogap phenomena in underdoped compounds and to the peculiar influence of hole carrier density on properties
of superconducting and normal states. Among the HTSCs discovered so far, La$_{2-x}$Sr$_{x}$CuO$_{4}$,
YBa$_{2}$Cu$_{3}$O$_{7-\delta}$ and Bi$_{2}$Sr$_{2}$CaCu$_{2}$O$_{8+\delta}$ are the mostly extensively studied
systems due to available single crystals over a large range of doping levels and transition temperatures.
La$_{2-x}$Sr$_{x}$CuO$_{4}$ has a structural phase transformation between the underdoped and overdoped
regimes.\cite{taka} Meanwhile, changing carrier density through cation substitution probably leads to local
distortion due to the size effect.\cite{attf}  Although the carrier density in YBa$_{2}$Cu$_{3}$O$_{7-\delta}$
can be tuned by oxygen content, the occurrence of chain oxygen sites in this material complicates interpretation
of much of the experimental data. Moreover, it is difficult to make even slightly overdoped samples in this
system, only the underdoped part of the phase diagram can be thoroughly studied. By comparison,
Bi$_{2}$Sr$_{2}$CaCu$_{2}$O$_{8+\delta}$ is a perfect candidate for studying its physical properties over a wide
range of the doping.

In order to get insight into the basic mechanism responsible for high-temperature superconductivity, one always
strives to observe the possible change of the superconducting transition temperature $T_{c}$ at fixed doping
level. Pressure has been realized to be an effective way to change $T_{c}$ for a compound. However, the effect
of pressure on $T_{c}$ is very complicated. The pressure derivative of $T_{c}$, $dT_{c}/dP$, can vary from
positive to negative, which strongly depends on the doping level. For most optimally doped
compounds,\cite{nmor,yama,sade,mori,jove,lgao,jove2} $T_{c}$ generally increases when pressure is applied at the
initial stage, passes through a saturation at some critical pressure level $P_{c}$, then decreases modestly at
high pressures. Bi$_{2}$Sr$_{2}$CaCu$_{2}$O$_{8+\delta}$ is the only known system where the constantly positive
$dT_{c}/dP$ is not sensitive to oxygen content,\cite{sieb,kubi,forr} which makes it an ideal system to
investigate the phase diagram between the saturated $T_{c}$ and critical pressure $P_{c}$ over an entire doping
regime. Early measurements on the 80$\sim$88 K samples\cite{alek,klot} have shown that for the sample with a
relatively high $T_{c}$ it is easy to reach the saturation at the beginning of the pressure range. It was also
shown that the $P_{c}$ shifts to higher pressure with additional oxygen content in the overdoped
side.\cite{alek,klot} It remains unclear whether the saturation in the $T_{c}$ versus $P$ curve is present for
samples away from optimal doping. Therefore, it is of interest to study the pressure dependence of $T_{c}$ as
well as the high-pressure phase diagram in Bi$_{2}$Sr$_{2}$CaCu$_{2}$O$_{8+\delta}$ over the entire doping
range.

Experiments carried out on a variety of HTSCs\cite{sade,loon} have revealed that $T_{c}$ is not a unique
function of pressure but depends strongly on the temperature at which the pressure is applied. For almost all
systems, pressure-induced relaxation effects become activated for temperatures above 200 K. The pressure
derivative $dT_{c}/dP$ can be increased tenfold if the pressure is changed at room temperature rather than at
low temperatures.\cite{whfi} Since the relaxation is believed to be frozen at low temperatures, the pressure
dependence of $T_{c}$ and the pressure derivative $dT_{c}/dP$ basically represent the intrinsic pressure effect.
Such an effect most directly reflects changes in the basic physical properties, including the pairing
interaction itself, and thus is most valuable for comparison with theory. It is therefore of interest to perform
detailed measurements of $T_{c}(P)$ under high pressure by changing pressure at low temperatures to elucidate
the intrinsic pressure effects.

In this paper we present the results of a high-pressure study of Bi$_{2}$Sr$_{2}$CaCu$_{2}$O$_{8+\delta}$ single
crystals in which the pressure is applied and measured at low temperatures. The characteristic nonmonotonic
pressure dependence of the superconducting transition temperature is found in our crystals ranging from heavily
underdoped to heavily overdoped samples. We obtain a new high-pressure phase diagram between the saturation of
$T_{c}$ and critical pressure $P_{c}$ at which $T_{c}$ is saturated. We demonstrate that the nonmonotonic
behavior originates from a competition between the hole carrier density and pairing interaction strength.

\section{EXPERIMENTAL}

\begin{figure}[tbp]
\begin{center}
\includegraphics[width=3.1 in]{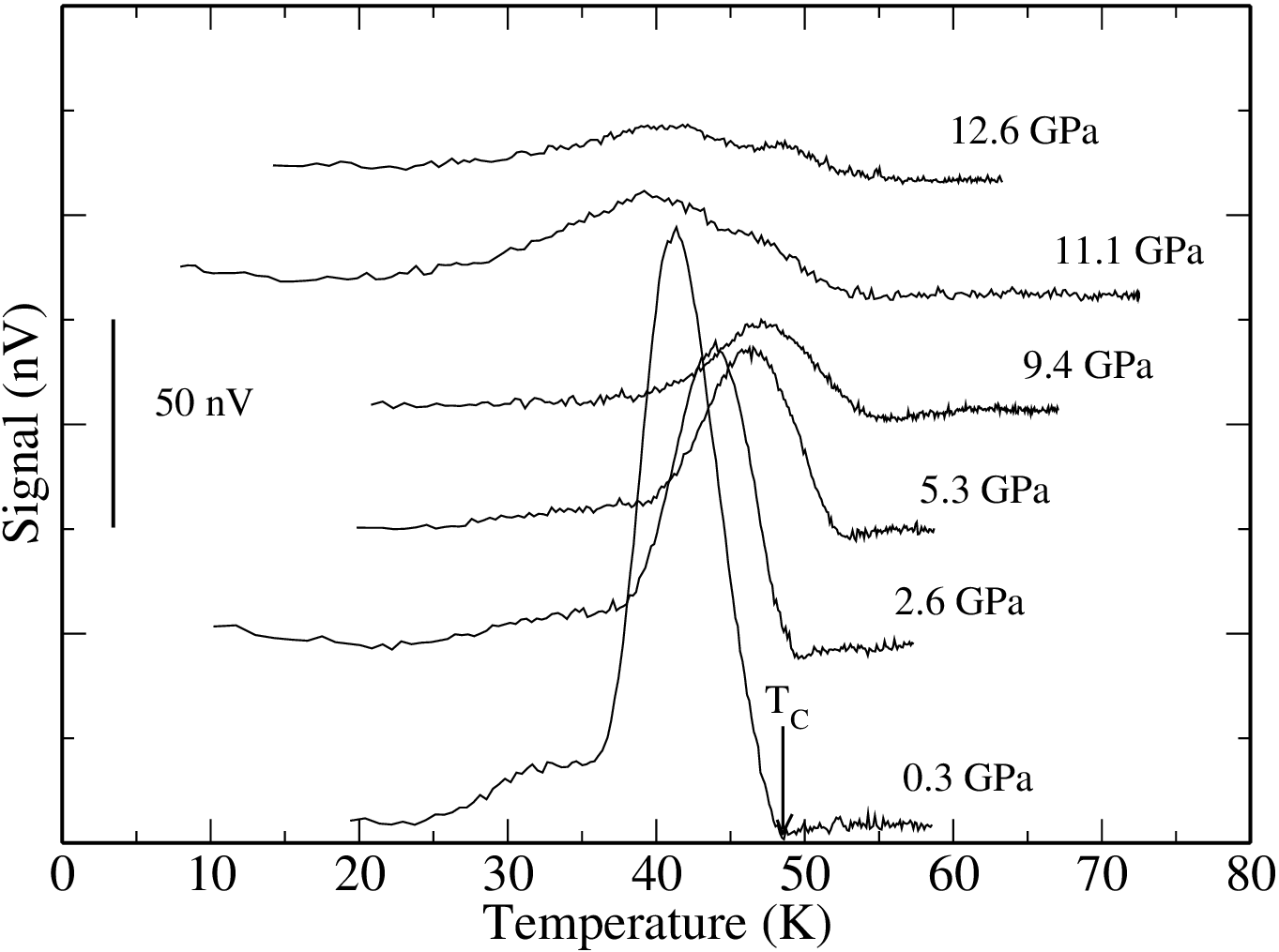}
\end{center}
\caption{ Magnetic susceptibility signal in nV vs. temperature for an underdoped Bi$_{2}$Sr$_{2}$CaCu$_{2}$O$_{8+\delta}$ 
crystal at various pressures. }
\end{figure}

Samples of single crystal Bi$_{2}$Sr$_{2}$CaCu$_{2}$O$_{8+\delta}$ were grown by a self-flux technique in a
strong thermal gradient to stabilize the direction of solidification using a stoichiometric ratio
(Bi:Sr:Ca:Cu=2:2:1:2) of cations, as described elsewhere.\cite{kend} Vacuum anneals are carried out in sealed
quartz tubes to obtain the underdoped crystals. Overdoping has been accomplished using stainless steel cells
sealed with samples immersed in liquid oxygen. Crystals used in the high-pressure studies are the highly
underdoped (``vacuum-annealed,'' $T_{c}=48$ K), air-annealed ($T_{c}=86$ K) and overdoped (``oxygen-annealed,''
$T_{c}=60$ K) samples. A 10-90\% transition width of 0.5 K in susceptibility of our samples is indicative of
very good quality and of high homogeneity in both cation and oxygen concentrations.\cite{kend}

We performed the measurements of $T_{c}$ under high pressures using a highly sensitive magnetic susceptibility
technique with diamond anvil cells.\cite{timo} This technique has been proven to be especially successful in the
discovery of superconductivity in compressed sulphur and lithium.\cite{stru} The technique is based on the
quenching of the superconductivity and suppression of the Meissner effect in the superconducting sample by an
external magnetic field. The magnetic susceptibility of the metallic parts of the high-pressure cells is
essentially independent of the external field. Therefore, the magnetic field applied to the sample inserted in
the diamond cell mainly affects the change of the signal coming from the sample. If we apply an alternating
low-frequency ($f=20$ Hz) magnetic field with an amplitude below 100 Oe, then the output signal changes twice
per period of the magnetic field, because the positive and negative magnetic fields act identically to destroy
the superconducting state. The resulting high-frequency output signal is modulated in amplitude at frequency
$2f$. This modulation is observed only in the narrow range of temperatures close to the temperature of
superconducting transition. The $T_{c}$ is then identified as the point where the signal goes to zero because of
the disappearance of the Meissner effect.

Samples are loaded in Mao-Bell cells made from hardened Be-Cu alloy. The gaskets are made of nonmagnetic Ni-Cr
alloy. The Bi$_{2}$Sr$_{2}$CaCu$_{2}$O$_{8+\delta}$ crystal (typical size $50\times 50\times 10$ $\mu$m$^{3}$)
together with ruby chips are placed in the sample chamber. NaCl is loaded into the gasket hole with diameter 200
$\mu$m to serve as pressure medium. The pressure is applied and measured at low temperatures and is gauged by
the $R$1 fluorescence line of ruby.\cite{mao} Each initial pressure run was performed as single run at low
temperature, pressure was changed immediately after the temperature scan was finished, and sample was taken back
to low temperature to start another temperature scan.

\section{RESULTS}

\begin{figure}[tbp]
\begin{center}
\includegraphics[width=\columnwidth]{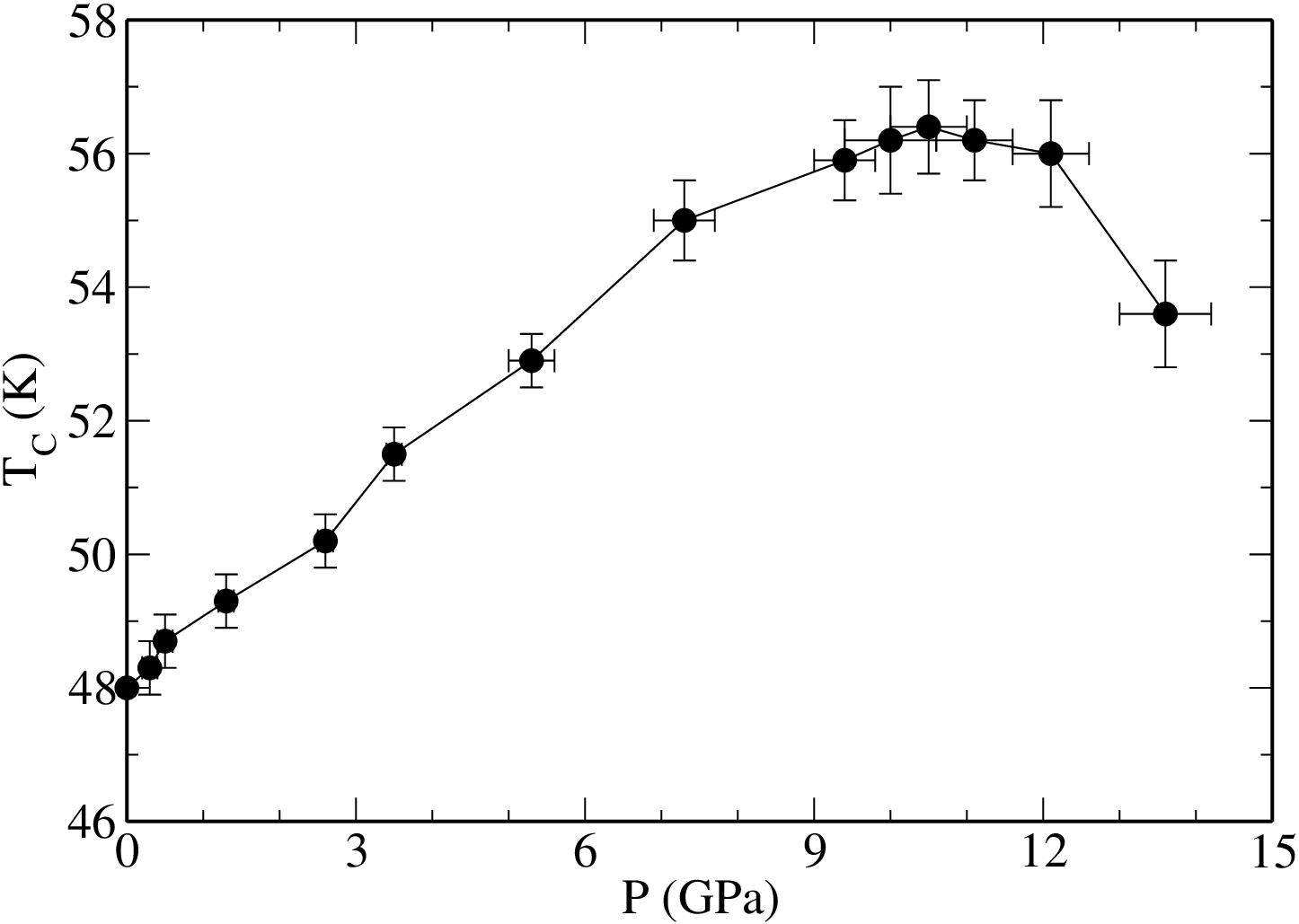}
\end{center}
\caption{ Dependence of the superconducting transition temperature $T_{c}$ on pressure in an underdoped
Bi$_{2}$Sr$_{2}$CaCu$_{2}$O$_{8+\delta}$ single crystal. }
\end{figure}

Figure 1 shows the representative temperature scans at different applied pressures for an underdoped
Bi$_{2}$Sr$_{2}$CaCu$_{2}$O$_{8+\delta}$ single crystal. Superconducting transition is identified as the
temperature where the signal goes to zero on the high-temperature side which is the point at which magnetic flux
completely enters the sample. The superconducting transition of 48 K is obtained at ambient pressure. It is
clear that at the pressure of 9.4 GPa the superconducting transition shifts to higher temperatures than at 1
atm, but it turns back beyond that pressure. The signal amplitude is suppressed by further increasing pressure.
Also, at high pressures the shape of the signal broadens. In our measurements, we apply pressure in the
increasing run and at low temperatures below 100 K. Therefore, we are confident for the existence of pressure
medium over the whole measured range.

In Fig. 2, we plot the obtained $T_{c}$ values in this underdoped Bi$_{2}$Sr$_{2}$CaCu$_{2}$O$_{8+\delta}$
sample as a function of pressure. As shown, $T_{c}$ is initially increased with applied pressure passing through
a saturation value at around $P_{c}=10.5$ GPa and then decreases at higher pressures. A positive initial
pressure derivative of 1.5 K/GPa is obviously consistent with the pure hydrostatic pressure measurements with
the helium gas system.\cite{sieb} The nonmonotonic pressure dependence of $T_{c}$ observed in the present
heavily underdoped Bi$_{2}$Sr$_{2}$CaCu$_{2}$O$_{8+\delta}$ is believed to be a common characteristic for all
underdoped $p$-type copper-oxides. This behavior has also been reported in the $\sim$80 K
Bi$_{2}$Sr$_{2}$CaCu$_{2}$O$_{8+\delta}$ samples with nearly optimal dopings.\cite{alek,klot} For an underdoped
Bi$_{2}$Sr$_{2}$CaCu$_{2}$O$_{8+\delta}$ single crystal with $T_{c}=54$ K which is close to our sample, Klotz
and Schilling\cite{klot} found an initial constancy of $T_{c}$ up to 1.6 GPa as well as the absence of the
saturation up to 6 GPa. This difference might be due to the difference of the samples and/or the difference in
the hydrostaticity of the pressure medium used.

\begin{figure}[tbp]
\begin{center}
\includegraphics[width=\columnwidth]{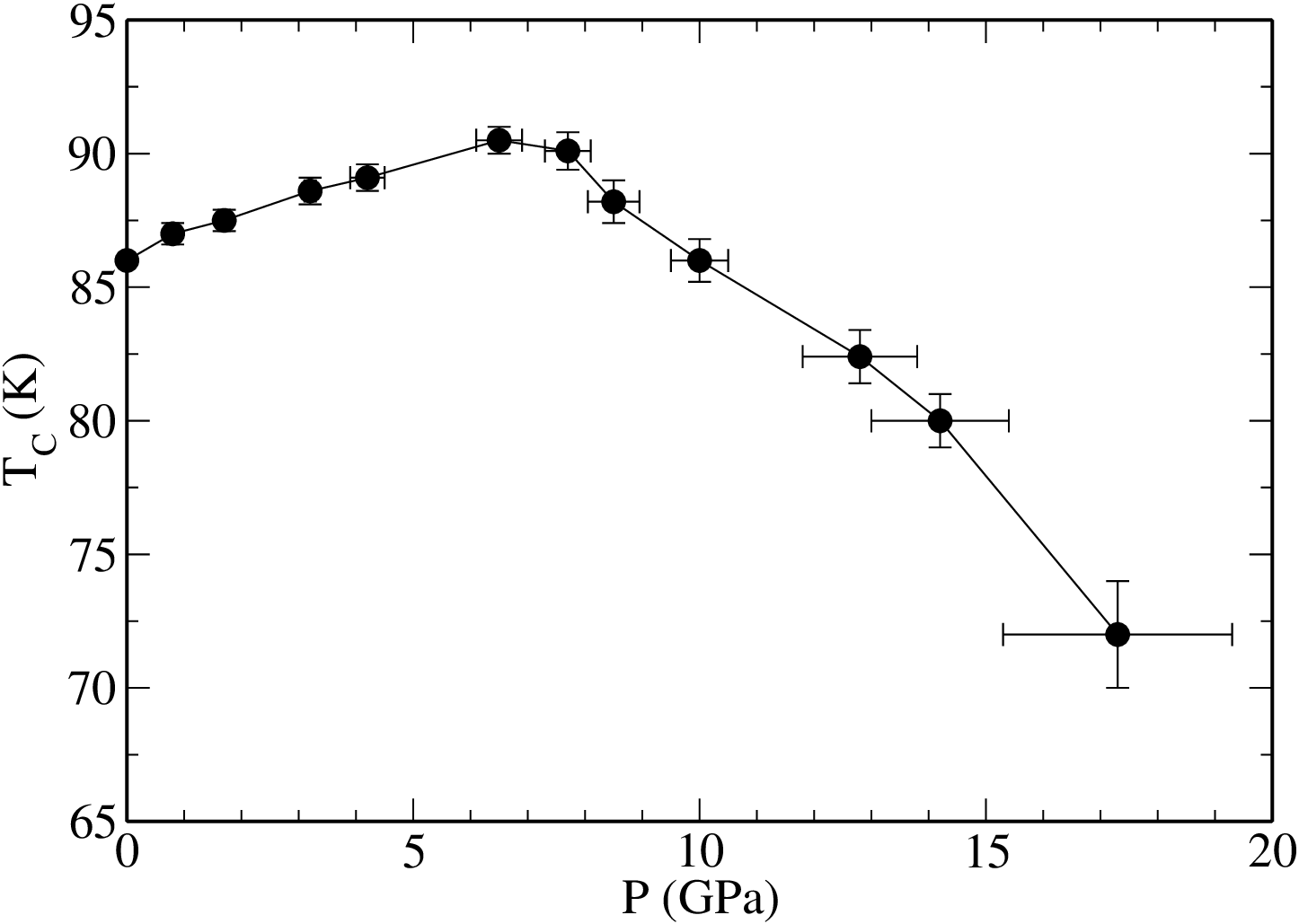}
\end{center}
\caption{ Shift of the superconducting transition temperature $T_{c}$ with applied pressure up to 18 GPa for a
Bi$_{2}$Sr$_{2}$CaCu$_{2}$O$_{8+\delta}$ single crystal which is nearly optimally doped but slightly overdoped. }
\end{figure}

The magnetic susceptibility curves for a nearly optimally doped but slightly overdoped
Bi$_{2}$Sr$_{2}$CaCu$_{2}$O$_{8+\delta}$ single crystal have a behavior similar to that in the underdoped
sample. The superconducting transition is at 86 K at ambient pressure and shifts upward when pressure is applied
at the beginning stage. Above 8.5 GPa, further pressure moves the transition to lower $T_{c}$ values. The
amplitude of the signal becomes weak with the applied pressure. It is hard to distinguish the signals from the
sample or background at pressures above 18 GPa.

Shown in Fig. 3, the pressure dependence of $T_{c}$ displays the characteristic behavior as is in other
optimally doped HTSCs.\cite{nmor,yama,sade,mori,jove,lgao} Similar nonmonotonic pressure dependence of
$T_{c}(P)$ with a saturation at a critical pressure near 2 GPa was reported in the sister bilayer
Tl$_{2}$Ba$_{2}$CaCu$_{2}$O$_{8+\delta}$ system at the optimal doping.\cite{mori} For the initial slope of the
$T_{c}$ versus $P$ curve in Fig. 3 we derive $dT_{c}/dP=1.3$ K/GPa similar to that measured in the helium gas
system.\cite{sieb} The saturation occurs at $P_{c}\approx 6.5$ GPa in our
Bi$_{2}$Sr$_{2}$CaCu$_{2}$O$_{8+\delta}$ sample at near optimal doping. This critical value is approximately the
same as that in the $T_{c}\sim 80$ K crystal of Alekseeva et al.,\cite{alek} but relatively higher than that for
the $T_{c}=88$ K sample of Klotz and Schilling.\cite{klot}

\begin{figure}[tbp]
\begin{center}
\includegraphics[width=\columnwidth]{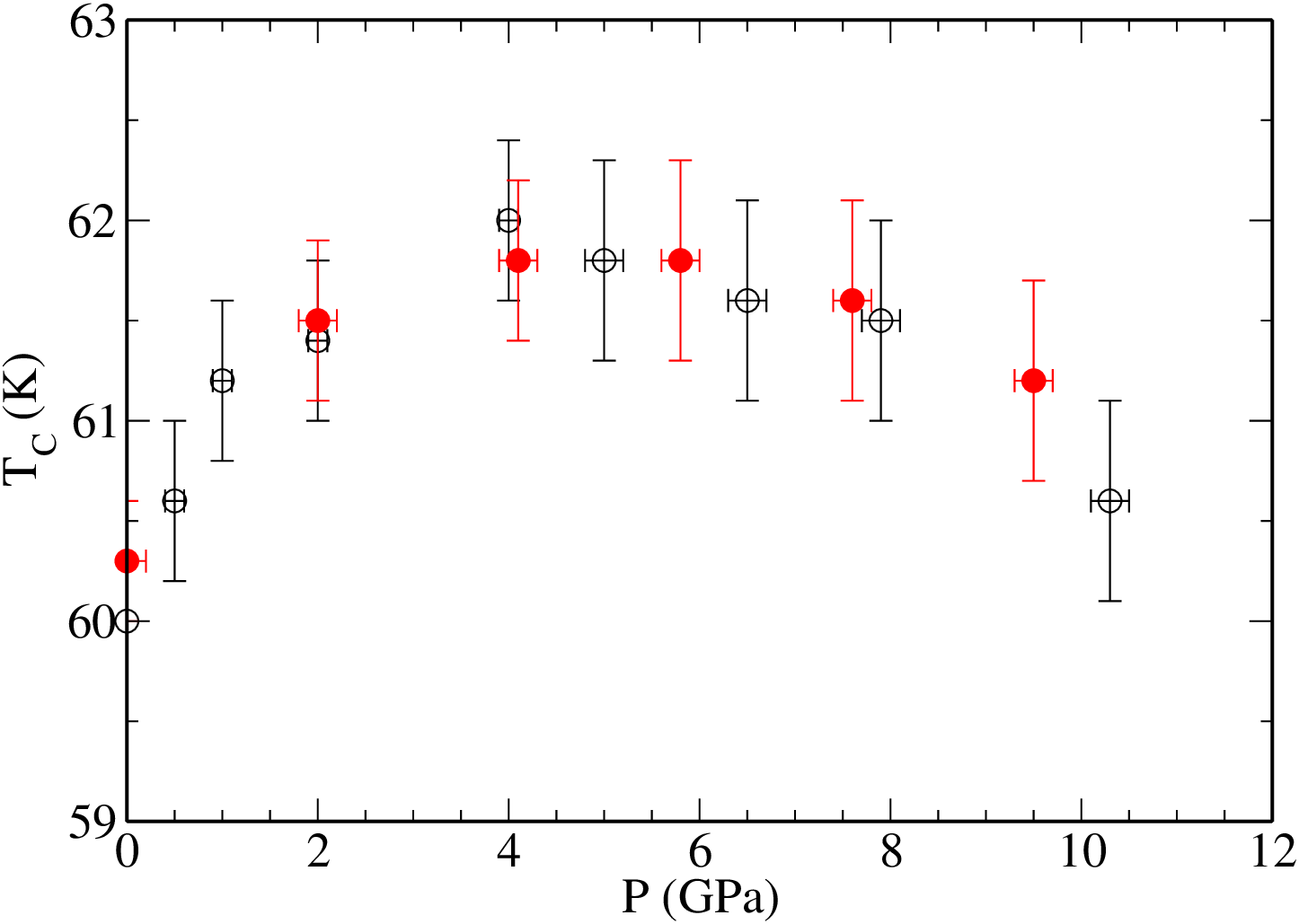}
\end{center}
\caption{ (Color online) Superconducting transition temperature $T_{c}$ for an overdoped
Bi$_{2}$Sr$_{2}$CaCu$_{2}$O$_{8+\delta}$ sample as a function of pressure. The open and solid circles represent
the measurements in the compressing and decompressing run, respectively. }
\end{figure} 

We have measured the temperature dependence of magnetic susceptibilities near $T_{c}$ for an overdoped
Bi$_{2}$Sr$_{2}$CaCu$_{2}$O$_{8+\delta}$ single crystal in both the compressing and decompressing runs. As
plotted in Fig. 4, the susceptibility curves are nearly identical, regardless of the pressure history. In the
decompressing run, $T_{c}$ almost returns to its original ambient pressure value upon full release of the
pressure. This is indicative of free relaxation in this sample.

From Fig. 4, it is found that as the applied pressure is increased, $T_{c}$ continues to increase until it
reaches a saturation value at near 4 GPa. Previous studies\cite{klot} on an overdoped
Bi$_{2}$Sr$_{2}$CaCu$_{2}$O$_{8+\delta}$ single crystal with a relatively high $T_{c}=80$ K have shown that the
$T_{c}$ also has a saturation value at near $P_{c}=4$ GPa. Compared to the maximum value of $T_{c}\approx 90$ K
in flux-grown Bi$_{2}$Sr$_{2}$CaCu$_{2}$O$_{8+\delta}$ at optimal doping, the present sample with a $T_{c}$ of
60 K is heavily overdoped. The occurrence of the saturation in the $T_{c}(P)$ curve in our sample is very
interesting, since it has never been reported in such a heavily overdoped cuprate where the significant change
of $T_{c}$ is tuned only through oxygen content rather than cation substitution. Considering the fact that
$T_{c}$ would decrease with further adding oxygen content in the overdoped regime, this result clearly indicates
that neither the pressure-induced charge transfer\cite{jorg} nor the pressure-induced oxygen doping\cite{yosh}
is solely responsible for the pressure effect on $T_{c}$. The observed saturation in the $T_{c}(P)$ curve also
suggests that another intrinsic variable independent of the hole carrier density is playing a significant role
in enhancing $T_{c}$ at even the heavily overdoped region.

\begin{figure}[tbp]
\begin{center}
\includegraphics[width=\columnwidth]{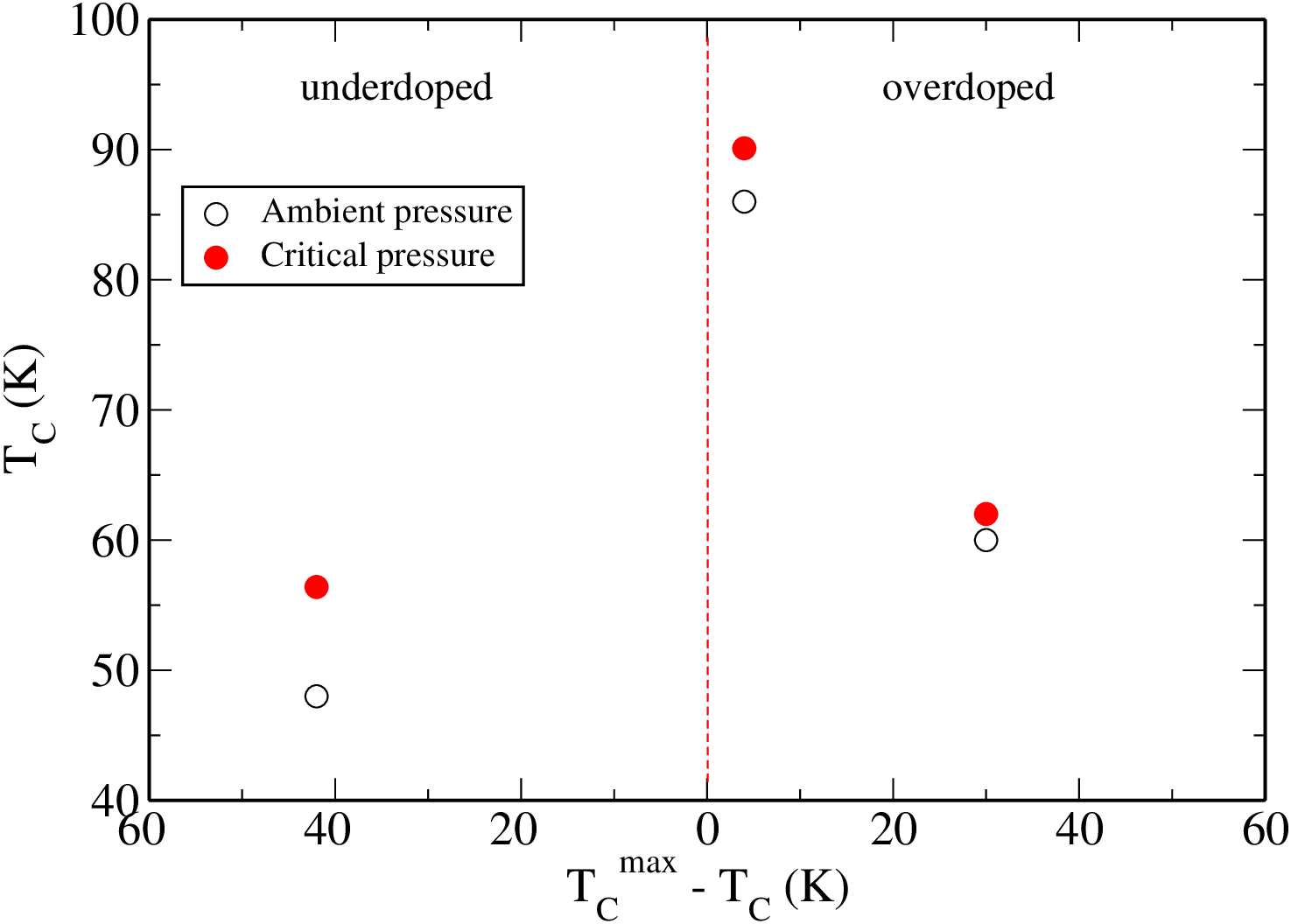}
\end{center}
\caption{ (Color online) Plot of the superconducting transition temperature $T_{c}$ at the ambient pressure (open 
circles) and critical pressure $P_{c}$ (solid circles) in flux-grown Bi$_{2}$Sr$_{2}$CaCu$_{2}$O$_{8+\delta}$ single 
crystals as a function of $T_{c}$ referenced to the maximum value of 90 K. }
\end{figure}

Among the results obtained in these experiments, the most important finding is that the saturation with larger
value than the ambient $T_{c}$ occurs in the Bi$_{2}$Sr$_{2}$CaCu$_{2}$O$_{8+\delta}$ samples over the entire
doping range from the heavily underdoped through the optimally doped to the heavily overdoped. Both the critical
pressure $P_{c}$ and difference between the $T_{c}$ and its saturated value decrease with the hole carrier
density created by adding oxygen into the Bi$_{2}$Sr$_{2}$CaCu$_{2}$O$_{8+\delta}$ system. This behavior is
similar to that reported for the La$_{2-x}$Sr$_{x}$CuO$_{4}$ system in the overdoped side\cite{nyam} where the
pressure-induced transition between the orthorhombic and tetragonal phases is accompanied when $T_{c}$ reaches
saturation.\cite{htak} Furthermore, we find the similar trend of $T_{c}(P_{c})$ in the present
Bi$_{2}$Sr$_{2}$CaCu$_{2}$O$_{8+\delta}$ samples at different doping levels as observed in other optimally doped
compounds.\cite{nmor,yama,sade,mori,jove,lgao,jove2} That is, the higher the $P_{c}$, the larger the saturation
value of the $T_{c}$ is. We thus obtain an extended phase diagram at both the ambient condition and critical
pressure, which is presented in Fig. 5. The saturation of $T_{c}$ under the critical pressure of a hole-doped
cuprate has been shown to cover the heavily underdoped to heavily overdoped regime. Despite extensive effort to
reaching the saturation at the critical pressure, such a high-pressure phase diagram over the whole doping
regime has not been established previously. The saturation $T_{c}(P_{c})$ was only found in either the
underdoped side\cite{sade} or the overdoped side.\cite{klot,nyam}

\section{DISCUSSION}

To understand the experimentally observed nonmonotonic behavior of $T_{c}(P)$ of these samples, we use a simple 
$d$-wave BCS formalism:
\begin{eqnarray}
1=\frac{1}{2N}\sum _{k}\frac{Vg^{2}(k)}{|\varepsilon _{k}-\mu|}\tanh(\frac{|\varepsilon _{k}-\mu|}{2k_{B}T_{c}})~~.
\end{eqnarray}
Where $g(k)=\cos k_x-\cos k_y$, $\varepsilon _{k}$ is the quasiparticle dispersion, $N$ is the number of $k$
vectors, $k_{B}$ is the Boltzmann constant, and $V$ is the in-plane pairing interaction strength. The constraint 
condition for the hole carrier density $n_{H}$ in conjunction with the chemical potential $\mu$ is given by
\begin{equation}
n_{H} = \frac{1}{2}-\frac{1}{2N}\sum _{k}\tanh(\frac{|\varepsilon _{k}-\mu|}{2k_{B}T_{c}})~~.
\end{equation}

The validity of this kind of $d$-wave BCS formalism in describing the superconducting state of HTSCs has been
confirmed by recent angle-resolved photoemission spectroscopy (ARPES)\cite{mats} and transport\cite{laty,prou}
measurements. As quasiparticle dispersion, we use\cite{dago} $\varepsilon _{k}=0.166\cos k_x\cos
k_{y}+0.046(\cos 2k_x+\cos 2k_y)$ eV, which corresponds to the hole dispersion in an antiferromagnetic
background at half-filling, but should be approximately valid also at finite $n_{H}$ as long as the
antiferromagnetic correlation length is not negligible. A characteristic feature of this dispersion is the near
flatness of the energy in the vicinity of $(\pi,0)$, which is in good agreement with ARPES results for
Bi$_{2}$Sr$_{2}$CaCu$_{2}$O$_{8+\delta}$.

The experimentally observed parabolic relation between $T_{c}$ and $n_{H}$ can be obtained by combining Eqs. (1)
and (2) for a given $V$. Taking the maximum $T_{c}$ of 90 K for flux-grown Bi$_{2}$Sr$_{2}$CaCu$_{2}$O$_{8+\delta}$, 
we deduce $V=0.03782$ eV. The resulting carrier densities $n_{H}$'s are then determined for the samples with various 
$T_{c}$ values.

The temperature dependences of both the in-plane resistivity $\rho_{ab}$ and Hall coefficient $R_{H}$ under high
pressures have been suspected of giving clues to the pressure effects on $T_{c}$. For optimally doped samples at
ambient pressure, the basic experimental trends are $\rho_{ab}\propto T$ and $R_{H}\propto 1/T$ in the high
temperature regime around room temperature. The quantity $R_{H}ez/v$ is inversely proportional to the hole
carrier density $n_{H}$, where $z$ is the number of Cu atoms per unit cell of volume $v$ and $e$ the carrier
charge. In the high-temperature $T$-linear region, the resistivity satisfies the simple form
$\rho_{ab}=(4\pi/\omega^{2}_{p})\tau^{-1}$, where $\omega_{p}$ is the plasma frequency and the scattering rate
$\tau^{-1}=2\pi\Lambda T$ with $\Lambda$ being the coupling strength between the charge carriers and some
excitations. We take the Drude spectral weight $\omega^{2}_{p} \sim n_{H}/m^{*}$ with $m^{*}$ being the carrier
effective mass. Systematic resistivity studies suggest that both the carrier scattering rate and coupling
strength are independent of doping and nearly same for a superconductor.\cite{tito} The pairing interaction
strength $V$ appeared in Eq. (1) should have the same physical meaning as $\Lambda$. Therefore, we have
$\rho_{ab}\propto (m^{*}/n_{H})VT$ and $R_{H}\propto v/(n_{H}ez)$, which should account for the resistivity and
Hall data at room temperature and under high pressures. Taking the pressure derivative of the quantities in the
expressions of $\rho_{ab}$ and $R_{H}$, we have the following relations:
\begin{eqnarray}
\frac{d\ln \rho_{ab}}{dP}&=& \frac{d\ln m^{*}}{dP} - \frac{d\ln n_{H}}{dP} + \frac{d\ln V}{dP}~~~, \\
\frac{d\ln R_{H}}{dP} &=& -\kappa _{v}- \frac{d\ln n_{H}}{dP}~~~.
\end{eqnarray}
Where the volume compressibility $\kappa _{v}\equiv 1/B_{0}=-d\ln v/dP$ and $B_{0}$ is the bulk modulus.

For the present Bi$_{2}$Sr$_{2}$CaCu$_{2}$O$_{8+\delta}$ system, we take $d\ln R_{H}/dP=-10.8\times 10^{-2}$
GPa$^{-1}$ and $\kappa _{v}=16.7\times 10^{-3}$ GPa$^{-1}$ from the Hall coefficient\cite{mura2} and neutron
diffraction\cite{gava} measurements, respectively. Substituting these values into Eq. (4), we obtain $d\ln
n_{H}/dP=9.1\times 10^{-2}$ GPa$^{-1}$. Studies of pressure effects on the irreversibility line\cite{raph} in a
nearly optimally doped Bi$_{2}$Sr$_{2}$CaCu$_{2}$O$_{8+\delta}$ reveal that the in-plane penetration depth
$\lambda _{ab}$ decreases from 220 to 205 nm when the applied pressure increases from 0 to 1.5 GPa, yielding
$d\ln \lambda _{ab}/dP=-4.55\times 10^{-2}$ GPa$^{-1}$. As it is known $\lambda _{ab}$ can be expressed in the
following relation: $\lambda^{2} _{ab} = m^{*}/n_{H}$. We then get $d\ln m^{*}/dP = 2d\ln \lambda _{ab}/dP +
d\ln n_{H}/dP$. According to this simple relation, we obtain the value of $d\ln m^{*}/dP$ of zero by using the
experimental data for $d\ln \lambda _{ab}/dP$ and $d\ln n_{H}/dP$. It is therefore reasonable to neglect the
pressure dependence of $m^{*}$. This indicates that the hole carrier density $n_{H}$ and pairing interaction
strength $V$ are the two main intrinsic pressure variables responsible for pressure effect besides the cell
volume $v$. This conclusion drawn from the normal-state transport properties agrees very well with the
assumption in interpreting the pressure effect in the Y-Ba-Cu-O systems.\cite{chen2}

High-pressure transport measurements give $d\ln \rho_{ab}/dP=-7.5\times 10^{-2}$ GPa$^{-1}$ for a nearly
optimally doped Bi$_{2}$Sr$_{2}$CaCu$_{2}$O$_{8+\delta}$ single crystal at 300 K.\cite{forr} Then we have $d\ln
V/dP=1.6\times 10^{-2}$ GPa$^{-1}$ by using Eq. (3). This result implies $d\ln V/dP \simeq \kappa_{v}$ in
Bi$_{2}$Sr$_{2}$CaCu$_{2}$O$_{8+\delta}$. Noting that the pressure-induced relative change of $V$ satisfies a
simple formalism $d\ln V/dP=(2\sim 4)\kappa _{a}$ in the Y-Ba-Cu-O systems,\cite{chen2} where $\kappa _{a}\equiv
-d\ln a/dP$ being the lattice compressibility along the $a$ axis. The relationship between the pairing
interaction strength and cell volume is then expressed by $\partial \ln V/\partial \ln v \simeq -1$. We would
like to mention that this probably universal relation does not rely on any model, at least in these two most
generally studied HTSCs.

Since $n_{H}$ and $V$ are two intrinsic pressure variables, the variation of $T_{c}$ with pressure is believed
to be due to the change of the competition between them. Assuming that the pressure dependence of both $n_{H}$
and $V$ are not as linear in pressure $P$, but in the relative volume decrease given by $1-v(P)/v_{0}$, we can
write $n_{H}(P)$ and $V(P)$ as
\begin{eqnarray}
n_{H}(P) &=& n_{H}(0)\left[1 + \frac{d\ln n_{H}}{dP}B_{0}\left(1-\frac{v(P)}{v_{0}}\right)\right]~~~,\nonumber \\
V(P) &=& V(0)\left[1 + \frac{d\ln V}{dP}B_{0}\left(1-\frac{v(P)}{v_{0}}\right)\right]~~~.
\end{eqnarray}
The pressure dependence of the relative volume can be well described by the first-order Murnaghan equation of
state $v(P)/v_{0}=(1+B_{0}^{\prime}P/B_{0})^{-1/B_{0}^{\prime}}$, where $B_{0}^{\prime}$ is the pressure
derivative of $B_{0}$ and $v_{0}$ the cell volume at ambient pressure. It has been shown that the construction
of the pressure dependence of the intrinsic pressure variable through the relative volume change always
reproduces an excellent agreement of the resulting curve with the experimental values.\cite{fiez,chen3} Olsen
$et$ $al.$\cite{olse} obtained $B_{0}^{\prime}=6.0$ for Bi-based cuprates by fitting their high-pressure X-ray
diffraction data up to 50 GPa. Their bulk modulus $B_{0}$ coincides well with that derived from the neutron
diffraction technique.\cite{gava}

The competition between the $n_{H}(P)$ and $V(P)$ provides natural explanation for the observed high-pressure 
phase diagram as shown in Fig. 5. $V(P)$ is a favorable factor for the increase of $T_{c}$ with pressure. 
While, $n_{H}(P)$ favors $T_{c}$ only for the compound in the underdoped regime, and it becomes a negative factor 
in the optimally dopant or the overdoped region. For the underdoped compound, $n_{H}(P)$ would be the negative
factor once the pressure-induced increase of $n_{H}$ passes through the optimal doping. The crucial parameter
$P_{c}$ is the crossover pressure where the factor(s) increasing $T_{c}$ is(are) equal to that(those) decreasing
$T_{c}$. A higher $P_{c}$ in our heavily underdoped sample arises from both the positive contribution to $T_{c}$
from $V(P)$ as well as the relatively long way for $n_{H}$ to go before passing through the optimal doping. The
nearly optimally doped compound has a modest $P_{c}$ due to the favorable factor $V(P)$ but the unfavorable
$n_{H}(P)$. Although $V(P)$ still factors the $T_{c}$ increase for the heavily overdoped material, the
pressure-induced increase of $n_{H}$ strongly suppresses such $T_{c}$ increase. As a result, there only exists a
smaller $P_{c}$ for the heavily overdoped material. Therefore, the extended high-pressure phase diagram can be
well understood on the basis of the competition of the two intrinsic pressure variables.

\begin{figure}[tbp]
\begin{center}
\includegraphics[width=\columnwidth]{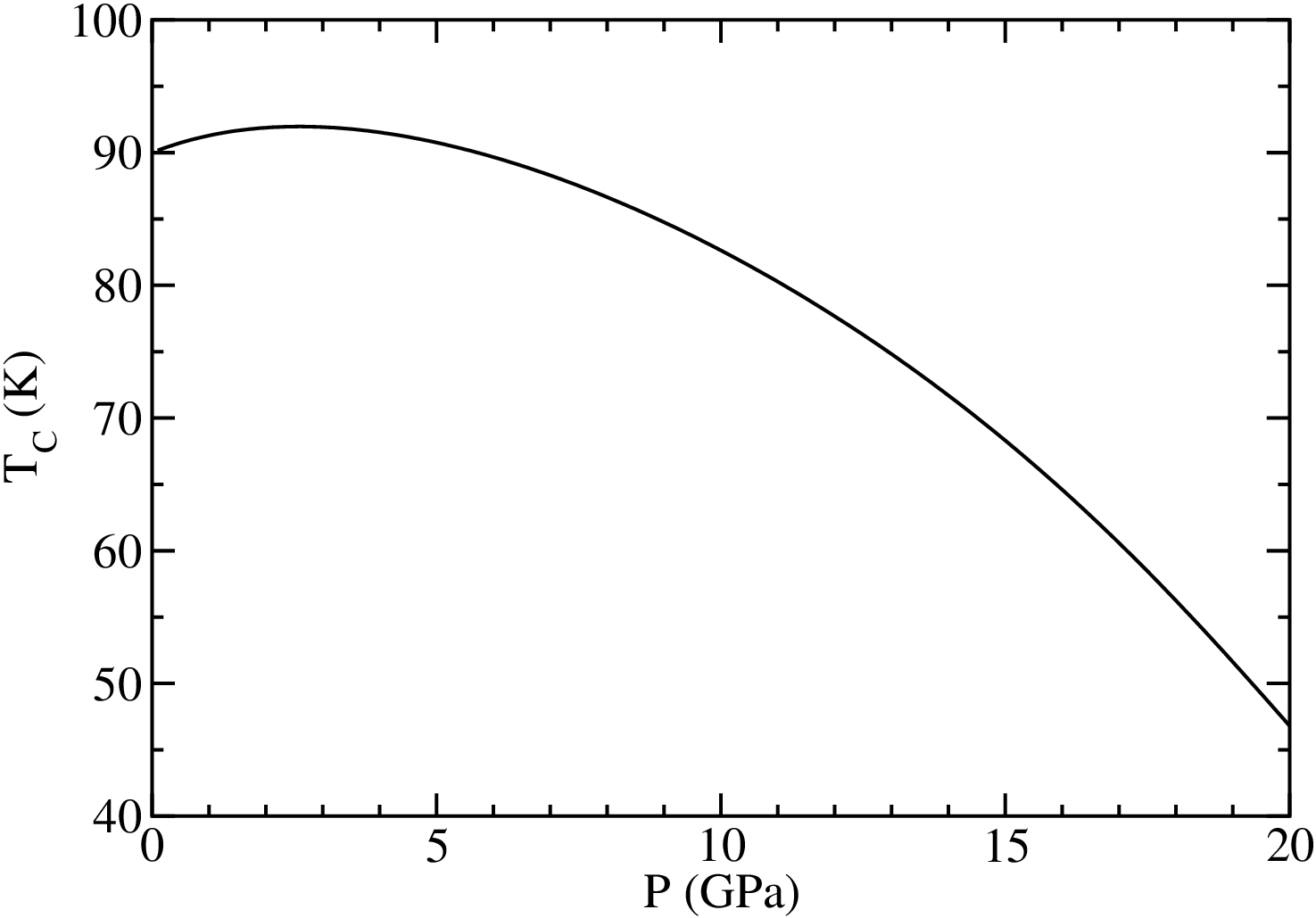}
\end{center}
\caption{ Calculated variation of the superconducting transition temperature $T_{c}$ with pressure for the optimally 
doped Bi$_{2}$Sr$_{2}$CaCu$_{2}$O$_{8+\delta}$.}
\end{figure}

Using the determined parameters $B_{0}$, $B_{0}^{\prime}$, $d\ln n_{H}/dP$, and $d\ln V/dP$, we are able to
evaluate the pressure dependence of $T_{c}$ in Bi$_{2}$Sr$_{2}$CaCu$_{2}$O$_{8+\delta}$ samples in terms of Eqs.
(1), (2), and (5). In Fig. 6 we plot the calculated variation of $T_{c}$ as a function of pressure up to 20 GPa 
for the optimally doped Bi$_{2}$Sr$_{2}$CaCu$_{2}$O$_{8+\delta}$ compound with $T_{c}=$90 K at ambient condition. 
As can be seen, as pressure is increased, $T_{c}$ increases initially until passing a saturation at critical 
pressure $P_{c}$ of 2.6 GPa, and at higher pressures $T_{c}$ decreases. This nonmonotonic behavior of $T_{c}(P)$ 
of the optimally doped compound is consistent with the measurements in our nearly optimally doped sample as well 
as the 88 K sample of Klotz and Schilling.\cite{klot} The theoretical value of the  initial $dT_{c}/dP$ is 1.69 
K/GPa for the optimally doped material, which is in excellent agreement with the reported $1.5\pm 0.2$ K/GPa from 
various measurements.\cite{sieb,kubi,forr} It is apparent that the competition between the hole carrier density and
pairing interaction strength indeed captures the essential physics of pressure effect on $T_{c}$.

\begin{figure}[tbp]
\begin{center}
\includegraphics[width=\columnwidth]{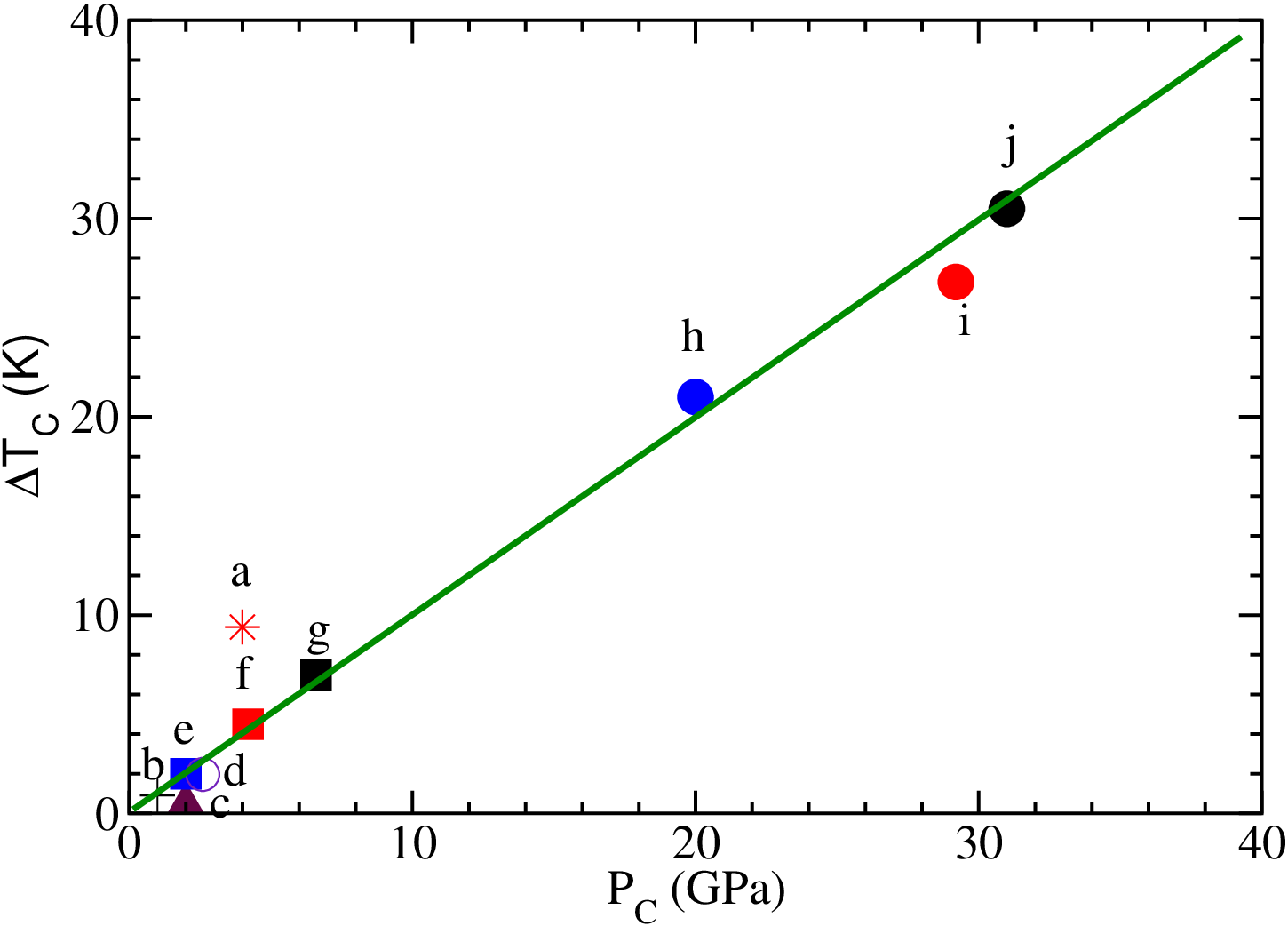}
\end{center}
\caption{ (Color online) Critical pressure $P_{c}$ dependence of the difference between the superconducting
transition temperatures ($\Delta T_{c}$) at the ambient pressure and critical pressure of optimally hole-doped 
HTSCs: a, La$_{2-x}$Sr$_{x}$CuO$_{4}$ (star, Ref. [\onlinecite{nmor}]); b, La$_{2-x}$Ca$_{1+x}$Cu$_{2}$O$_{6}$ 
(plus, Ref. [\onlinecite{yama}]); c, YBa$_{2}$Cu$_{3}$O$_{7-\delta}$ (triangle, Ref. [\onlinecite{sade}]); d,
Bi$_{2}$Sr$_{2}$CaCu$_{2}$O$_{8+\delta}$ (open circle, this work); e, Tl$_{2}$Ba$_{2}$CaCu$_{2}$O$_{8+\delta}$ 
(Ref. [\onlinecite{mori}]); f, Tl$_{2}$Ba$_{2}$Ca$_{2}$Cu$_{3}$O$_{10+\delta}$ (Ref. [\onlinecite{jove}]); g,
Tl$_{2}$Ba$_{2}$Ca$_{3}$Cu$_{4}$O$_{12+\delta}$ (Ref. [\onlinecite{jove}]); h, HgBa$_{2}$CuO$_{4+\delta}$ (Ref.
[\onlinecite{lgao}]); i, HgBa$_{2}$CaCu$_{2}$O$_{6+\delta}$ (Ref. [\onlinecite{lgao}]); j,
HgBa$_{2}$Ca$_{2}$Cu$_{3}$O$_{8+\delta}$ (Ref. [\onlinecite{lgao}]). The line is the guide to the eyes. }
\end{figure}

Figure 7 shows the critical pressure $P_{c}$ dependence of the difference between the superconducting transition
temperatures $\Delta T_{c}$ at the ambient pressure and critical pressure in the optimally doped
Bi$_{2}$Sr$_{2}$CaCu$_{2}$O$_{8+\delta}$ with those in some other optimally doped
compounds.\cite{nmor,yama,sade,mori,jove,lgao} It can be seen that there exists the same trend of increasing the
$\Delta T_{c}$ value with the increase of $P_{c}$ in these optimally doped compounds. $\Delta T_{c}$ increases
under $P_{c}$ at the rate of near 1 K/GPa. It is indicated that any optimally doped compound has a saturation
value of the maximum $T_{c}$ which can be reached by applied pressure at critical $P_{c}$ value. Moreover, it is
confirmed that pressure is a most effective way to enhance $T_{c}$ in optimally hole-doped cuprates. These
results also suggest that the enhancement of the maximum $T_{c}$  with pressure in the optimally doped HTSCs is
not due to the increase of the pressure-induced hole carrier density.

Since the above determined parameters are available only for the optimally or nearly optimally doped materials,
we can not calculate the change of $T_{c}$ with pressure for various dopings by using the same parameters,
especially for $d\ln n_{H}/dP$ and $d\ln V/dP$. There are not enough high-pressure transport data available in
obtaining the information on these parameters at the moment. However, we can estimate the pressure-induced
change of carrier density $n_{H}$, if we assume that $d\ln V/dP$ is independent of oxygen content just as the
interaction strength $V$ is. Substituting the experimentally determined $dT_{c}/dP$ of 1.5 K/GPa for
Bi$_{2}$Sr$_{2}$CaCu$_{2}$O$_{8+\delta}$,\cite{sieb} we get the $d\ln n_{H}/dP$ values of 5.1$\times 10^{-3}$,
6.4$\times 10^{-3}$, and 2.5$\times 10^{-3}$ GPa$^{-1}$ for the underdoped, nearly optimally doped, and
overdoped samples with $T_{c}=48$, 86, and 60 K, respectively. For comparison, the high-pressure transport data
provide the $d\ln n_{H}/dP$ value of 9.1$\times 10^{-2}$ GPa$^{-1}$ for the optimally doped compound with
$T_{c}=90$ K. This means that the optimally doped compound has the largest amount of the pressued-induced
increase of the hole carrier density among the samples with different doping level. However, as discussed above,
the pressure effect on $T_{c}$ in this material is still dominated by the pairing interaction strength.

\section{CONCLUSION}

We have measured the pressure dependence of the superconducting transition temperature $T_{c}$ in flux-grown
Bi$_{2}$Sr$_{2}$CaCu$_{2}$O$_{8+\delta}$ single crystals by using a highly sensitive magnetic susceptibility
technique with diamond anvil cells. In order to avoid the possible contribution from the pressure relaxation
effect, we apply and measure pressure at low temperatures. The nonmonotonic pressure behavior of $T_{c}$ is
observed for all the samples, from the heavily underdoped through the nearly optimally doped to the heavily
overdoped regime. We therefore obtain a high-pressure phase diagram between the saturation value of $T_{c}$ 
and critical pressure $P_{c}$. Theoretical calculation based on a $d$-wave BCS formalism reproduces well this 
nonmonotonic behavior in terms of the high-pressure transport data. We have demonstrated that the pressure 
dependence of $T_{c}$ in Bi$_{2}$Sr$_{2}$CaCu$_{2}$O$_{8+\delta}$ is the result of the competition between 
the hole carrier density and pairing interaction strength. The generally observed $T_{c}(P_{c})$ in our samples 
indicates that the pairing interaction strength is mainly responsible for the pressure effect in this system.

\begin{acknowledgments}
The authors are grateful to Profs. C. W. Chu and J. S. Schilling for valuable comments and helpful discussions.
This work at Carnegie was supported by the U.S. Department of Energy under awards DEFG02-02ER4595 and
DEFC03-03NA00144. C. K. acknowledges support from the U.S. Office of Naval Research.
\end{acknowledgments}


\begin{thebibliography}{99}

\bibitem{taka} H. Takagi, R. J. Cava, M. Marezio, B. Batlogg, J. J. Krajewski, W. F. Peck, Jr., P. Bordet, and
D. E. Cox, Phys. Rev. Lett. {\bf 68}, 3777 (1992).

\bibitem{attf} J. P. Attfield, A. L. Kharlanov, and J. A. McAllister, Nature (London) {\bf 394}, 157 (1998).

\bibitem{mori} N. M\^{o}ri, H. Takahashi, Y. Shimakawa, T. Manako, and Y. Kubo, J. Phys. Soc. Jpn. {\bf 59},
L3839 (1990).

\bibitem{nmor} N. M\^{o}ri, C. Murayama, H. Takahashi, H. Kaneko, K. Kawabata, Y. Iye, S. Uchida, H. Takagi, Y.
Tokura, Y. Kubo, H. Sasakura, and K. Yamaya, Physica C {\bf 185-189}, 40 (1991).

\bibitem{sade} S. Sadewasser, J. S. Schilling, A. P. Paulikas, and B. W. Veal, Phys. Rev. B {\bf 61}, 741
(2000).

\bibitem{yama} Y. Yamada, K. Kinoshita, T. Matsumoto, F. Izumi, and T. Yamada, Physica C {\bf 185-189}, 1299
(1991).

\bibitem{jove} D. Tristan Jover, R. J. Wijngaarden, R. Griessen, E. M. Haines, J. T. Tallon, and R. S. Liu,
Phys. Rev. B {\bf 54}, 10175 (1996).

\bibitem{lgao} L. Gao, Y. Y. Xue, F. Chen, Q. Xiong, R. L. Meng, D. Ramirez, C. W. Chu, J. H. Eggert, and H. K.
Mao, Phys. Rev. B {\bf 50}, 4260 (1994).

\bibitem{jove2} D. Tristan Jover, H. Wilhelm, R. J. Wijngaarden, and R. S. Liu, Phys. Rev. B {\bf 55}, 11832 (1997).

\bibitem{sieb} R. Sieburger, P. M\"{u}ller, and J. S. Schilling, Physica C {\bf 181}, 335 (1991).

\bibitem{kubi} R. Kubiak, K. Westerholt, G. Pelka, H. Bach, and Y. Khan, Physica C {\bf 166}, 523 (1990).

\bibitem{forr} L. Forr\'{o}, V. Ilakovac, and B. Keszei, Phys. Rev. B {\bf 41}, 9551 (1990).

\bibitem{alek} E. A. Alekseeva, I. V. Berman, N. B. Brandt, A. A. Zhukov, I. L. Romashkina, and V. I. Sidorov,
JETP Lett. {\bf 51}, 467 (1990) [Pis'ma Zh. Eksp. Teor. Fiz. {\bf 51}, 411 (1990)].

\bibitem{klot} S. Klotz and J. S. Schilling, Physica C {\bf 209}, 499 (1993).

\bibitem{loon} see, R. Sieburger and J. S. Schilling, Physica C {\bf 173}, 423 (1991); A.-K. Klehe, C. Looney,
J. S. Schilling, H. Takahashi, N. M\^{o}ri, Y. Shimakawa, Y. Kubo, T. Manako, S. Doyle, and A. M. Hermann,
$ibid.$ {\bf 257}, 105 (1996); S. Sadewasser, Y. Wang, J. S. Schilling, H. Zheng, A. P. Paulikas, and B. W.
Veal, Phys. Rev. B {\bf 56}, 14168 (1997); C. Looney, J. S. Schilling, and Y. Shimakawa, Physica C {\bf 297},
239 (1998); S. Sadewasser, J. S. Schilling, J. L. Wagner, O. Chmaissem, J. D. Jorgensen, D. G. Hinks, and B.
Dabrowski, Phys. Rev. B {\bf 60}, 9827 (1999); S. Sadewasser, J. S. Schilling, A. Knizhnik, G. M. Reisner, and
Y. Eckstein, Eur. Phys. J. B {\bf 15}, 15 (2000).

\bibitem{whfi} W. H. Fietz, R. Quenzel, H. A. Ludwig, K. Grube, S. I. Schlachter, F. W. Hornung, T. Wolf, A.
Erb, M. Kl\"{a}ser, and G. M\"{u}ller-Vogt, Physica C {\bf 270}, 258 (1996).

\bibitem{kend} C. Kendziora, R. J. Kelley, E. Skelton, and M. Onellion, Physica C {\bf257}, 74 (1996).

\bibitem{timo} Y. A. Timofeev, V. V. Struzhkin, R. J. Hemley, H. K. Mao, and E. A. Gregoryanz, Rev. Sci. Instr.
{\bf 73}, 371 (2002).

\bibitem{stru} V. V. Struzhkin, R. J. Hemley, H. K. Mao, and Y. A. Timofeev, Nature (London) {\bf 390}, 382 (1997);
V. V. Struzhkin, M. I. Eremets, W. Gan, H. K. Mao, and R. J. Hemley, Science {\bf 298}, 1213 (2002).

\bibitem{mao} H. K. Mao, J. Xu, and P. M. Bell, J. Geophys. Res. {\bf 91}, 4673 (1986).

\bibitem{jorg} J. D. Jorgensen, S. Pei, P. Lightfoot, D. G. Hinks, B. W. Veal, B. Dabrowski, A. P. Paulikas, R.
Kleb, and I. D. Brown, Physica C {\bf 171}, 93 (1990).

\bibitem{yosh} K. Yoshida, A. I. Rykov, S. Tajima, and I. Terasaki, Phys. Rev. B {\bf 60}, R15035 (1999).

\bibitem{nyam} N. Yamada and M. Ido, Physica C {\bf 203}, 240 (1992).

\bibitem{htak} H. Takahashi, H. Shaked, B. A. Hunter, P. G. Radaelli, R. L. Hitterman, D. G. Hinks, and J. D.
Jorgensen, Phys. Rev. B {\bf 50}, 3221 (1994).

\bibitem{mats} H. Matsui, T. Sato, T. Takahashi, S.-C. Wang, H.-B. Yang, H. Ding, T. Fujii, T. Watanabe, and A.
Matsuda, Phys. Rev. Lett. {\bf 90}, 217002 (2003).

\bibitem{laty} Yu. I. Latyshev, T. Yamashita, L. N. Bulaevskii, M. J. Graf, A. V. Balatsky, and M. P. Maley,
Phys. Rev. Lett. {\bf 82}, 5345 (1999).

\bibitem{prou} C. Proust, E. Boaknin, R. W. Hill, L. Taillefer, and A. P. Mackenzie, Phys. Rev. Lett. {\bf 89},
147003 (2002).

\bibitem{dago} E. Dagotto, A. Nazarenko, and A. Moreo, Phys. Rev. Lett. {\bf 74}, 310 (1995).

\bibitem{tito} T. Ito, K. Takenaka, and S. Uchida, Phys. Rev. Lett. {\bf 70}, 3995 (1993).

\bibitem{mura2} C. Murayama, Y. Iye, T. Enomoto, N. M\^{o}ri, Y. Yamada, T. Matsumoto, Y. Kuto, Y. Shimakawa,
and T. Manako, Physica C {\bf 183}, 277 (1991).

\bibitem{gava} J.-R. Gavarri, O. Monnereau, G. Vacquier, C. Carel, and C. Vettier, Physica C {\bf 172}, 213 (1990).

\bibitem{raph} M. P. Raphael, M. E. Reeves, E. F. Skelton, and C. Kendziora, Phys. Rev. Lett. {\bf 84}, 1587 (2000).

\bibitem{chen2} X. J. Chen, H. Q. Lin, and C. D. Gong, Phys. Rev. Lett. {\bf 85}, 2180 (2000).

\bibitem{fiez} W. H. Fietz, F. W. Hornung, K. Grube, S. I. Schlachter, T. Wolf, B. Obst, and P. Schweiss, J. Low
Temp. Phys. {\bf 117}, 915 (1999).

\bibitem{chen3} X. J. Chen, H. Zhang, and H.-U. Habermeier, Phys. Rev. B {\bf 65}, 144514 (2002).

\bibitem{olse} J. S. Olsen, S. Steenstrup, L. Gerward, and B. Sundqvist, Phys. Scr. {\bf 44}, 211 (1991).

\end{thebibliography}
\end{document}